\documentclass[twocolumn,eqsecnum,aps]{revtex4}

\usepackage[dvips]{graphicx} % PRC
\usepackage{color}           % PRC % for color  usage: \textcolor{red}{text}
\newcommand{\beq}{\begin{eqnarray}}
\newcommand{\eeq}{\end{eqnarray}}

\newcommand{\nc}{\newcommand}           % new command myo
\nc{\vc}[1]     {\mbox{\boldmath $#1$}} % boldmath(vector) myo
\begin{document}

%\parindent=10pt
%%%%%%%%%%%%%%%%%%%%%%%%%%%%%%%%%%%%%%%%%%%%%%%%%%%%%%%%%%%%%%%%

\title{Structure of neutron-rich He $\Lambda$ hypernuclei
using the cluster orbital shell model}

\author {T. Myo$^{1,2}$}

\affiliation{$^1$General Education, Faculty of Engineering, Osaka Institute of Technology, Osaka 535-8585, Japan}
\affiliation{$^2$Research Center for Nuclear Physics (RCNP), Osaka University, Ibaraki 567-0047, Japan}

\author{E.\ Hiyama$^{3,4}$}

\affiliation{$^3$Department of Physics, Tohoku University, Sendai, 980-8578, Japan}

\affiliation{$^3$RIKEN, Nishina Center,  Wako, Saitama, 351-0198,Japan}

%\date{\today}
%
%%%%%%%%%%%%%%%%%%%%%%%%%%%%


\begin{abstract}
We calculated the energy spectra of the neutron-rich He $\Lambda$ hypernuclei with $A=6$ to 9
within the framework of an $\alpha + \Lambda +Xn$ ($X=1$--4) cluster model using the cluster orbital shell model.
The employed constituent particles reproduce their observed properties.
For resonant states of core nuclei such as $^5$He, $^6$He, and $^7$He, the complex scaling method is
employed to obtain energies and decay widths. The calculated ground states of $^6_{\Lambda}$He
and $^7_{\Lambda}$He are in good agreement with published data. The energy levels of $^8_{\Lambda}$He and
$^9_{\Lambda}$He are predicted.  In $^9_{\Lambda}$He,
we find one deeply bound state and two excited resonant states, which are proposed to be produced at 
the Japan proton accelerator research complex (J-PARC)
by the double-charge-exchange reaction $(\pi^-, K^+)$ using a $^9$Be target.
\end{abstract}

\maketitle

%===============================================================
\section{Introduction}
%===============================================================

In unstable, non-strange nuclear physics, nuclei near the proton and neutron drip-lines
produce interesting phenomena involving a neutron halo exotic structure.
For instance,  He isotopes with $A$=3 to 10 have been observed.
From these observations we find that the ground state of $^6$He has a halo structure,
the ground state of $^8$He is bound (the edge of the drip-line of the He isotopes is $^8$He), 
and the ground states of $^9$He and $^{10}$He are unbound. 

In hypernuclei, it is also interesting to search for the neutron and proton drip lines.
Because there is no Pauli principle acting between
nucleons and a $\Lambda$ hyperon, inserting a $\Lambda$ into a nucleus
gives rise to more bound states due to the attractive nature of 
the $\Lambda N$ interaction.
Thus, if a $\Lambda$ particle is added to a neutron-rich nucleus with
weakly bound or resonant neutrons, the resulting
hypenucleus will be more stable against neutron decay.
(This phenomenon is called a ``glue-like'' role.)
Thanks to this effect, there is a chance to produce a hypernuclear
neutron (proton) halo state, if the core nucleus has a weakly unbound (resonant) state
with an appropriate energy above the particle decay threshold.
In fact, when a $\Lambda$ is added to He isotopes,
the binding energies of all the resulting He hypernuclei ($^4_{\Lambda}$He, 
$^5_{\Lambda}$He, $^6_{\Lambda}$He, and $^7_{\Lambda}$He)
become larger than the binding energies of their core nuclei, as one should anticipate.
In particular, $^7_{\Lambda}$He was observed using the the $(e,e'K^+)$ reaction
at Jefferson Laboratory (JLAB) in 2013, and a $\Lambda$ separation energy of $B_{\Lambda}=5.68 \pm ({\rm stat.})
\pm 0.25 ( {\rm sys.}) $ MeV was reported \cite{Nakamura}.
After this discovery, the experiment  was repeated
with five times more statistics, and the observed ground state $B_{\Lambda}$ 
was reported to be $5.55 \pm 0.10 ({\rm stat.}) \pm 0.11 ({\rm sys.})$ \cite{Gogami}.
The first excited state of $^7_{\Lambda}$He was also
observed with $B_{\Lambda}$ is $3.65  \pm 0.20 ({\rm stat.}) \pm 0.11 ({\rm sys.})$\cite{Gogami}.
Theoretically, we have studied these hypernuclei, 
$^5_{\Lambda}$He, $^6_{\Lambda}$He and $^7_{\Lambda}$He, using  
$\alpha \Lambda$, $\alpha \Lambda N$ ,and $\alpha \Lambda nn$ two-body,
three-body, and four-body cluster models, respectively \cite{Hiyama96,Hiyama2009,
Hiyama2015}.
According to Ref.\cite{Hiyama96}, although the core nucleus $^5$He is unbound,
due to the attractive nature of the $\Lambda N$ interaction 
(resulting in the glue-like role of a $\Lambda$ particle), 
the ground state of $^6_{\Lambda}$He is a weakly bound state with respect to the 
$^5_{\Lambda}{\rm He}+n$ threshold.  As a result, this hypernucleus exhibits an 
extended valence neutron density, or neutron halo.
It is well known that the ground state of $^6$He is a halo state.
Thus, the $^7_{\Lambda}$He hypernucleus is more bound with the addition of a
$\Lambda$, and there is no neutron halo.
However, with the addition of a $\Lambda$, the $2^+$ excited state of $^6$He 
becomes a weakly bound state in $^7_{\Lambda}$He.  

Another important goal in the study of neutron-rich $\Lambda$ hypernuclei is
to extract information about $\Lambda N$-$\Sigma N$ coupling and
to obtain how much $\Sigma$ probability is contained in hypernuclei.
In fact, to obtain information on  $\Lambda N$-$\Sigma N$ coupling,
intensive theoretical calculations of $^4_{\Lambda}$H and
$^4_{\Lambda}$He has been performed \cite{Hiyama2001,Nogga,Nemura}
using realistic hyperon-nucleon ($YN$) interactions.
In Ref. \cite{Hiyama2001},
$^4_{\Lambda}$H and $^4_{\Lambda}$He were studied
taking into account $\Lambda N$-$\Sigma N$ coupling explicitly; it was 
concluded that  $\Lambda N$-$\Sigma N$ coupling was important
to make these $A=4$ $\Lambda$ hypernuclei bound.
However,  because of the ambiguity in realistic $YN$ potential models,
understanding $\Lambda N$-$\Sigma N$ coupling remains an open question.
To obtain further information on  $\Lambda N$-$\Sigma N$ coupling,
neutron-rich $\Lambda$ hypernuclei such as the He isotopes are well suited. 
Because $\Lambda$ hypernuclei of the He isotopes with $A=4$ to 7
have been observed,
systematic study of these $\Lambda$ hypernuclei would be
very useful to obtain information on $\Lambda N$-$\Sigma N$ coupling.
For the He core nuclear isotopes, it is  well known that $^8$He is the end 
of the bound systems.
By the injecting a $\Lambda$ particle into $^8$He, the resulting $^9_{\Lambda}$He 
would be more deeply bound. 
If $^9_{\Lambda}$He can be observed, it would be also helpful for the study of
$\Lambda N$-$\Sigma N$ coupling.

So far, $^9_{\Lambda}$He has not been observed. However,
plans are to produce this $\Lambda$ hypernucleus at the Japan proton accelerator research complex (J-PARC)
using the double-charge-exchange reaction $(\pi^-, K^+)$ on a
$^9$Be target.  
Therefore, it is timely to study He isotope $\Lambda$ hypernuclei.
In Ref.\cite{wirth} He isotope $\Lambda$ hypernuclei together with their
corresponding core nuclei have been calculated systematically within a
no-core shell model.  The authors employed $NN$, $NNN$, and $YN$ 
potentials based upon chiral effective field theory \cite{Machleidt,Navratil,Haidenbauer}.
Because the authors did not focus on reproducing the observed binding energies 
of the hypernuclei and the core nuclei,  it was
difficult to predict the energy spectra of $^9_{\Lambda}$He.

The aim of the present paper is to predict the energy spectra of $^9_{\Lambda}$He
within the framework of an $\alpha +4n+\Lambda$ six-body cluster model
using the cluster orbital shell model (COSM), which has been applied to
He isotope nuclei for bound states as well as resonant states \cite{myo21}.
By combining the COSM with the complex scaling method, we
could obtain decay widths for resonant states in these nuclei.
To demonstrate the validity of this model approach,
we calculate the energy spectra of $^5_{\Lambda}$He,
$^6_{\Lambda}$He, $^7_{\Lambda}$He, and  $^8_{\Lambda}$He and those of the
corresponding core nuclei.
For He isotope $\Lambda$ hypernuclei, we show our theoretical binding energies 
are in good agreement with the observed data using an effective 
single channel $\Lambda N$ 
interaction, where the $\Lambda N -\Sigma N$ couple-channel potential
is renomalized into a single-channel $\Lambda N$ potential, because
there is no realistic $YN$ interaction 
which reproduces the observed data of light $\Lambda$ hypernuclei. 

The paper is organized as follows:
In Section II, the method and interactions are outlined. In Section III,
the results are presented and discussed.  Finally, in Section IV, we summarize
our conclusions.

%===============================================================
\section{Model and Interaction}
%===============================================================

\subsection{Cluster orbital shell model}

\begin{figure}[b]
\begin{center}
\includegraphics[width=4.5cm,clip]{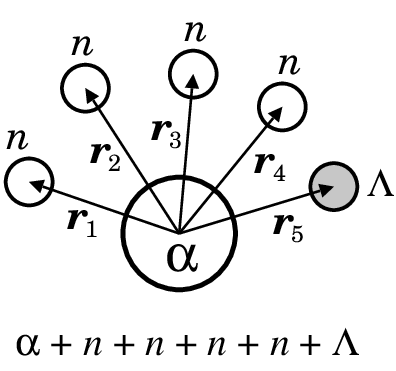}
\caption{Coordinate system of $\alpha$+$n$+$n$+$n$+$n$+$\Lambda$ in the COSM.}
\label{fig:COSM}
\end{center}
\end{figure}

We describe the neutron-rich  $\Lambda$ He isotopes assuming an $\alpha$ cluster represents the strongly bound $^4$He central core.
We explain the six-body cluster model of $^9_\Lambda$He with $\alpha$+$n$+$n$+$n$+$n$+$\Lambda$ in the cluster orbital shell model (COSM) \cite{suzuki88,masui06}.
The relative coordinates of the five valence particles around the $\alpha$ cluster are $\{\vc{r}_i\}$ with $i=1,\ldots,5$ as shown in Fig.~\ref{fig:COSM}.
The coordinate $\vc{r}_5$ is that of the $\Lambda$ particle.
The Hamiltonian is  
\begin{eqnarray}
	H
&=&	t_\alpha+ \sum_{i=1}^{A_v} t_i + t_\Lambda - T_G + \sum_{i=1}^{A_v} v^{\alpha n}_i + \sum_{i<j}^{A_v} v^{nn}_{ij}
	\nonumber\\
&+&    v^{\alpha \Lambda} + \sum_{i=1}^{A_v} v^{n \Lambda}_i
    \\
&=&	\sum_{i=1}^{A_v} \left( \frac{\vc{p}^2_i}{2\mu_{\alpha n}} + v^{\alpha n}_i \right) 
      + \sum_{i<j}^{A_v} \left( \frac{\vc{p}_i\cdot \vc{p}_j}{A_\alpha m} + v^{nn}_{ij} \right) 
    \nonumber\\
&+&	\frac{\vc{p}^2_\Lambda}{2\mu_{\alpha \Lambda}} + v^{\alpha \Lambda} 
      + \sum_{i=1}^{A_v} \left( \frac{\vc{p}_i\cdot \vc{p}_\Lambda}{A_\alpha m} + v^{n \Lambda}_{i} \right)
      \label{eq:Ham}
\end{eqnarray}
where $A_v$ and $A_\alpha$ are the number of valence neutrons and a mass number of the $\alpha$ particle, respectively. 
The mass $m$ is a nucleon mass.
The kinetic energy operators $t_\alpha$, $t_i$, $t_\Lambda$, and $T_G$ are those of the $\alpha$, neutron, $\Lambda$, and center-of-mass part, respectively.
The operator $\vc{p}_i$ ($\vc{p}_\Lambda$) is the relative momentum between the $\alpha$ and a valence neutron ($\Lambda$)
with the reduced mass $\mu_{\alpha n}$ ($\mu_{\alpha \Lambda}$).
The nucleon part of the Hamitonian is the same as used in the previous studies~\cite{myo14,myo21,myo22}.
The $\alpha$--neutron interaction $v^{\alpha n}$ is given by the microscopic Kanada-Kaneko-Nagata-Nomoto potential \cite{kanada79}.
For the neutron-neutron interaction $v^{nn}$, we use the Minnesota central potential \cite{tang78}
and slightly modify the potential to fit the observed two-neutron separation energy of $^6$He.
For the $\Lambda$-neutron interaction $v^{\Lambda n}$, we adopt the phenomenological single-range Gaussian potential, ORG \cite{motoba85}.
For the $\alpha$-$\Lambda$ interaction, we adopt the folding potential of $v^{\Lambda n}$ using the density of the $\alpha$ particle
with the $s$-wave configuration, which reproduces the $B_\Lambda=3.12$ MeV observed in $^5_\Lambda$He. \cite{motoba85,hiyama97}.

%%%%%%%%%%%%%%%%%%%%%%%%%%%%%
%%%%%%%%%%%%%%%%%%%%%%%%%%%%%
 
The total wave function with spin $J$ of the He isotopes with $\Lambda$ is given by the superposition of the configurations $\Psi^J_c$ in the COSM as
\begin{eqnarray}
    \Psi^J
&=& \sum_c C^J_c \Psi^J_c,
    \qquad
    \Psi^J_c
~=~ \prod_{i=1}^{A_v} a^\dagger_{p_i}a^\dagger_{p_\Lambda}|0\rangle, 
    \label{WF0}
\end{eqnarray}
where the vacuum $|0\rangle$ indicates the $\alpha$ cluster.
We adopt the $(0s)^4$ configuration of the harmonic oscillator wave function for the $\alpha$ cluster.
The range parameter of the $0s$ orbit is 1.4 fm to reproduce the charge radius of $^4$He.
The operator $a^\dagger_{p_i}$ creates a single-particle state $p_i$ of a valence neutron with the coordinate $\vc{r}_i$ in a $jj$-coupling scheme.
For the $\Lambda$, the creation operator is $a^\dagger_{p_\Lambda}$ with the coordinate of $\vc{r}_5$ in Fig. \ref{fig:COSM}.
The index $c$ is the set of $\{p_i\}$ and $p_\Lambda$ for valence particles and specifies the configuration $\Psi^J_c$.
For valence neutron states, we impose the orthogonality condition, in which the relative $0s$ orbit is removed in the state $\phi_p$.

We expand the radial part of $\phi_p(\vc{r})$ with a finite number of Gaussian functions $u(\vc{r},b)$ for each single-particle state:
\begin{eqnarray}
    \phi_p(\vc{r})
&=& \sum_{q=1}^{N_{\ell j}} d^q_p\ u_{\ell j}(\vc{r},b_{\ell j}^q)\, ,
    \label{spo}
    \\
    u_{\ell j}(\vc{r},b_{\ell j}^q)
&=& r^{\ell} e^{-(r/b_{\ell j}^q)^2/2}\, [Y_{\ell}(\hat{\vc{r}}),\chi^\sigma_{1/2}]_{j}\, ,
    \label{Gauss}
	\\
    \langle \phi_p | \phi_{p'} \rangle 
&=& \delta_{p,p'}
~=~ \delta_{n,n'}\, \delta_{\ell,\ell'}\, \delta_{j,j'}.
    \label{Gauss2}
\end{eqnarray}
The label $q$ specifies the Gaussian range parameter $b_{\ell j}^q$ with $q=1,\ldots, N_{\ell j}$,
where $N_{\ell j}$ is a basis number.
The parameters $\{b_{\ell j}^q\}$ are given in the geometric progression \cite{Hiyama03}.
The coefficients $\{d^q_p\}$ in Eq.~(\ref{spo}) are determined using the orthogonality condition of the basis states $\{\phi_p\}$ in Eq.~(\ref{Gauss2}),
where $n$ is the label to distinguish the radial part of $\phi_p(\vc{r})$.
The number $N_{\ell j}$ is determined to get the convergence of the solutions and we use $N_{\ell j}=10$ at most
with the range of $b_{\ell j}^q$ from 0.3 fm to around 30 fm.
In the COSM,, each of the configurations $\Phi^J_c$ in Eq.~(\ref{WF0}) becomes the product of the single-particle basis states $\phi_p$.
Using the Gaussian basis states with various range parameters, we can describe the weak-binding state of the multivalence particles in the COSM,
and also treat the resonances by applying complex scaling.

For the single-particle states $\phi_p$, we include the orbital angular momenta $\ell\le 2$.
We use 173.7 MeV for the repulsive strength of the Minnesota potential $v^{nn}$ instead of the original 200 MeV
to reproduce 0.975 MeV for the two-neutron separation energy of $^6$He.
This condition is the same as used in the previous works \cite{myo10,myo21,myo22}
and reproduces the energy spectra of the He isotopes and their mirror proton-rich nuclei.
For $^9_\Lambda$He, due to the large model space of the configuration mixing, we limit the spin and parity of the core nucleus of $^8$He to be the $0^+$ state,
because the excitation energy of the first excited state of $^8$He ($2^+$) is 3.4 MeV, which is rather high in the present discussion of the binding energy of the $\Lambda$ in $^9_\Lambda$He.

We combine the available configurations $\Psi^J_c$ with the amplitude of $C_c^J$ in Eq.~(\ref{WF0}).
The Hamiltonian matrix elements are calculated analytically using the COSM configurations with Gaussian basis states.
One finally solves the eigenvalue problem of the Hamiltonian matrix as
\begin{eqnarray}
  \sum_{c'}\langle \Psi^J_c |H|  \Psi^J_{c'} \rangle\, C^J_{c'} &=& E^J C_{c}^J  ,
  \label{eq:eigen}
  \\
  \sum_c (C^J_{c})^2&=&1.
\end{eqnarray}
We obtain the amplitudes $\{C^J_c\}$ and the energy eigenvalues $E^J$ of the He isotopes with $\Lambda$,
measured from the six-body threshold energy of $^4$He+$n$+$n$+$n$+$n$+$\Lambda$.

%%%%%%%%%%%%%%%%%%%%%%%%%%%%%%%%%%%%%%%%%%%%%
\subsection{Complex scaling method}

We explain the complex scaling method to treat resonances and continuum states in the many-body systems \cite{ho83,moiseyev98,aoyama06,myo14,myo20}.
The resonance is defined to be the eigenstate having the complex eigenenergy with the outgoing boundary condition
and the continuum states are orthogonal to the resonances.
In the complex scaling, all of the particle coordinates $\{\vc{r}_i\}$ as shown in Fig. \ref{fig:COSM}, are transformed using a common scaling angle $\theta$ as
\begin{eqnarray}
\vc{r}_i \to \vc{r}_i\, e^{ i\theta},\qquad
\vc{p}_i \to \vc{p}_i\, e^{-i\theta} .
\end{eqnarray}
The Hamiltonian in Eq.~(\ref{eq:Ham}) is transformed into the complex-scaled Hamiltonian $H_\theta$, and the complex-scaled Schr\"odinger equation is written as
\begin{eqnarray}
	H_\theta\Psi^J_\theta
&=&     E^J_\theta \Psi^J_\theta .
	\label{eq:eigen}
        \\
    \Psi^J_\theta
&=& \sum_c C^J_{c,\theta} \Psi^J_c.
    \label{WF_CSM}
\end{eqnarray}
The eigenstates $\Psi^J_\theta$ are obtained by solving the eigenvalue problem in Eq.~(\ref{eq:eigen}).
The total wave function $\Psi^J_\theta$ has a $\theta$ dependence, which is included in the expansion coefficients $C_{c,\theta}^J$ in Eq.~(\ref{WF_CSM}).
We obtain the energy eigenvalues $E^J_\theta$ of the bound and unbound states on a complex energy plane, which are governed by the ABC theorem \cite{ABC}.
From the ABC theorem, the asymptotic boundary condition of resonances is transformed to the damping behavior. 
Due to the boundary condition of the resonances with the complex scaling, we can use the same numerical method as the one to obtain the bound states in the calculation of resonances.
In the complex scaling, the Riemann branch cuts are commonly rotated down by $2\theta$ in the complex energy plane, 
in which each of the branch cuts starts from the corresponding threshold energy.
On the other hand, the energy eigenvalues of the bound and resonant states are independent of $\theta$ from the ABC theorem.
From these properties, one can identify the resonances with complex energy eigenvalues as $E_r-i\Gamma/2$, 
where $E_r$ and $\Gamma$ are the resonance energies and the decay widths, respectively. 
The scaling angle $\theta$ is determined in each resonance to give the stationary point of the energy eigenvalue on the complex energy plane.

For the resonance wave function, it is proved that its asymptotic condition becomes the damping form if $2\theta > |\arg(E_{\rm R})|$ in the complex energy plane \cite{ABC}.
Hence the resonance wave functions can be expanded in terms of the $L^2$ basis functions because of the damping boundary condition,
and the amplitudes of resonances are normalized with the condition of $\sum_{c} \big(C^J_{c,\theta}\big)^2=1$.
It is noted that the Hermitian product is not adopted due to the bi-orthogonal property of the complex-energy eigenstates including resonances \cite{berggren68,ho83,moiseyev98}.

\section{Results}

\begin{figure}[htb]
\begin{center}
\includegraphics[width=8.0cm,clip]{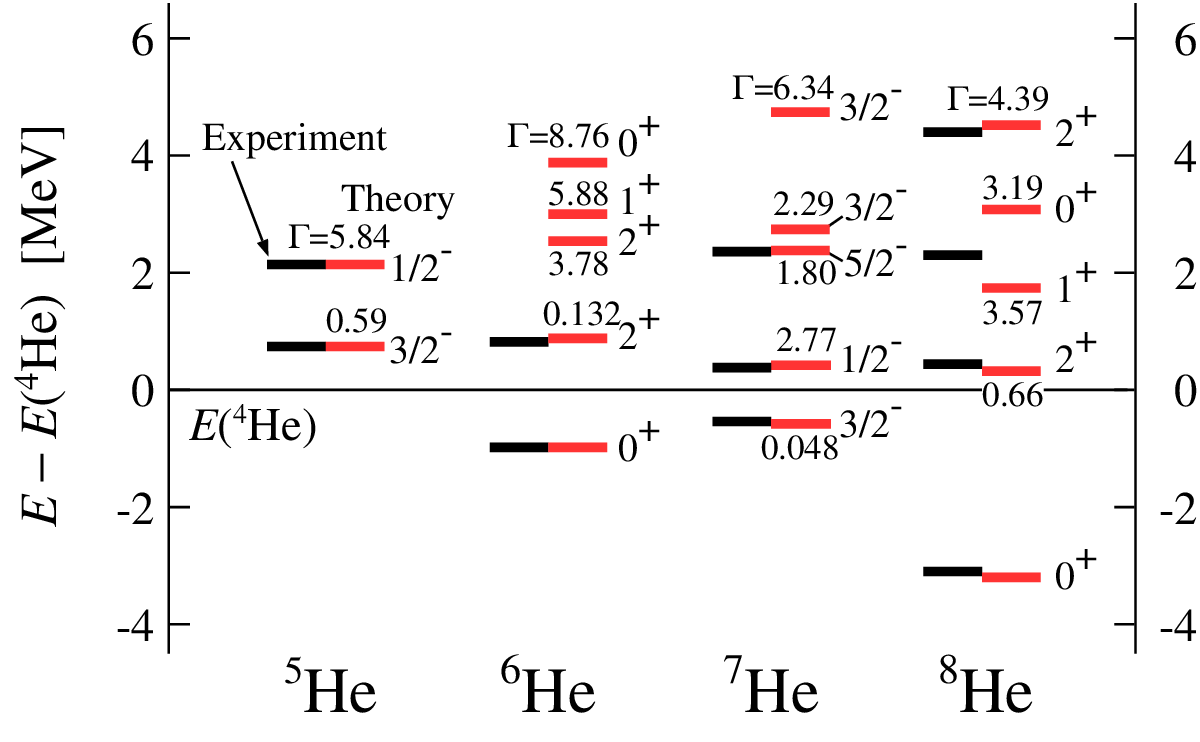}
\caption{The calculated energy spectra of the He isotope nuclei
together with experimental data. Black lines are experimental data and
red line are calculated values. The experimental data are taken 
in Refs.\cite{Tilley,Tilley-he,Skaza,Golovkov,Holl}. Numbers above the red lines
are values of decay widths. The energies are measured
with respect to $\alpha +Xn$ ($X=1$--4) threshold.}
\label{fig:he}
\end{center}
\end{figure}

First, let us discuss the energy spectra of He isotope nuclei
as shown in Fig.2. The energies are weakly bound states and resonant states.
Regarding resonant states, we utilize the complex scaling method to obtain energies and decay widths
as explained in Sec.II.B.
We see that the calculated energy spectra (red lines) of
$^5$He, $^6$He and $^7$He are in good agreement with the
observed data (black lines) well.
In the case of $^8$He, the observed binding energy is $-3.1$ MeV, while,
the calculated one is $-3.21$ MeV, which is consistent with the data.
Also, we calculate other excited states with decay widths for $^8$He.
We see that the calculated excited states are not inconsistent with
 the observed states. 
To show the reliability of our model, we also illustrate the
rms radii of the bound states among the He isotopes, namely, 
ground states in $^6$He and $^8$He in Fig.3.
We see that the calculated rms radii are in good agreement with the data.
This means that we obtain reliable wavefunctions 
for $^6$He and $^8$He.

\begin{figure}[t]
\begin{center}
\includegraphics[width=8.0cm,clip]{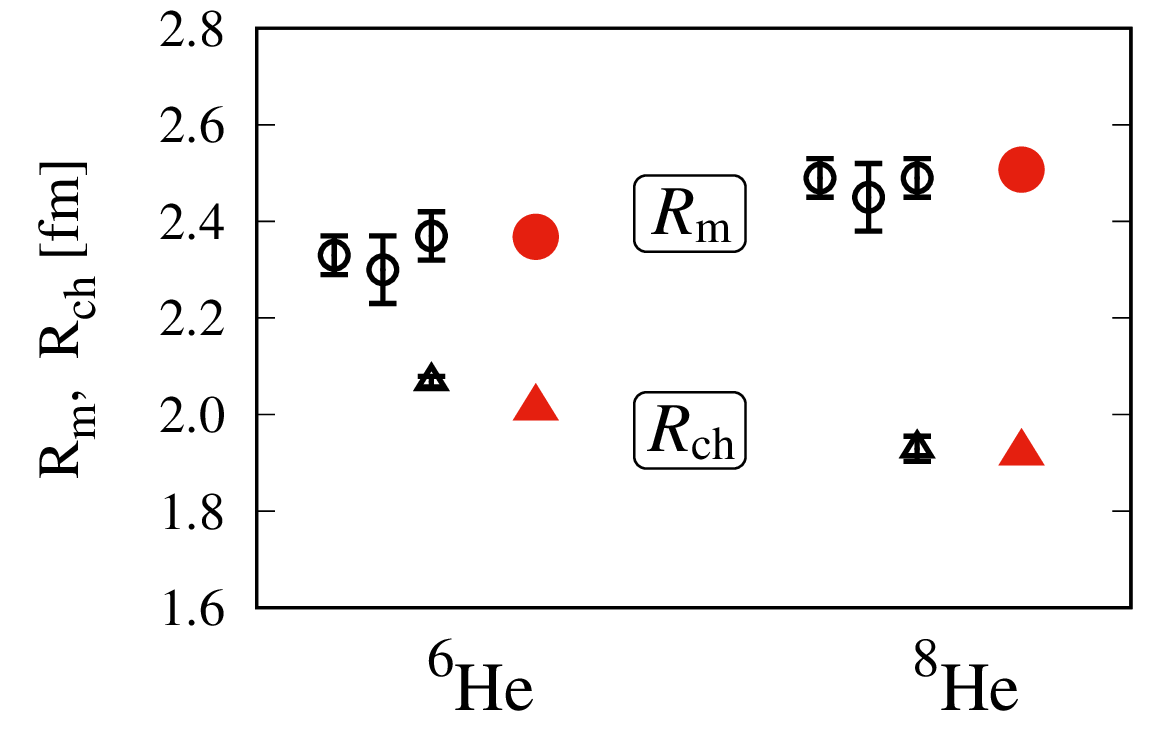} % new
\caption{The calculated rms radii of $^6$He and $^8$He for matter ($R_{\rm m}$, circle) 
and charge ($R_{\rm ch}$, triangle) with red solid symbols together with
experimental data. The data are taken from Refs.\cite{tanihata92,alkazov97,kiselev05, mueller07,brodeur12}. }
\label{fig:rms}
\end{center}
\end{figure}

\begin{figure}[htb]
\begin{center}
\includegraphics[width=8.0cm,clip]{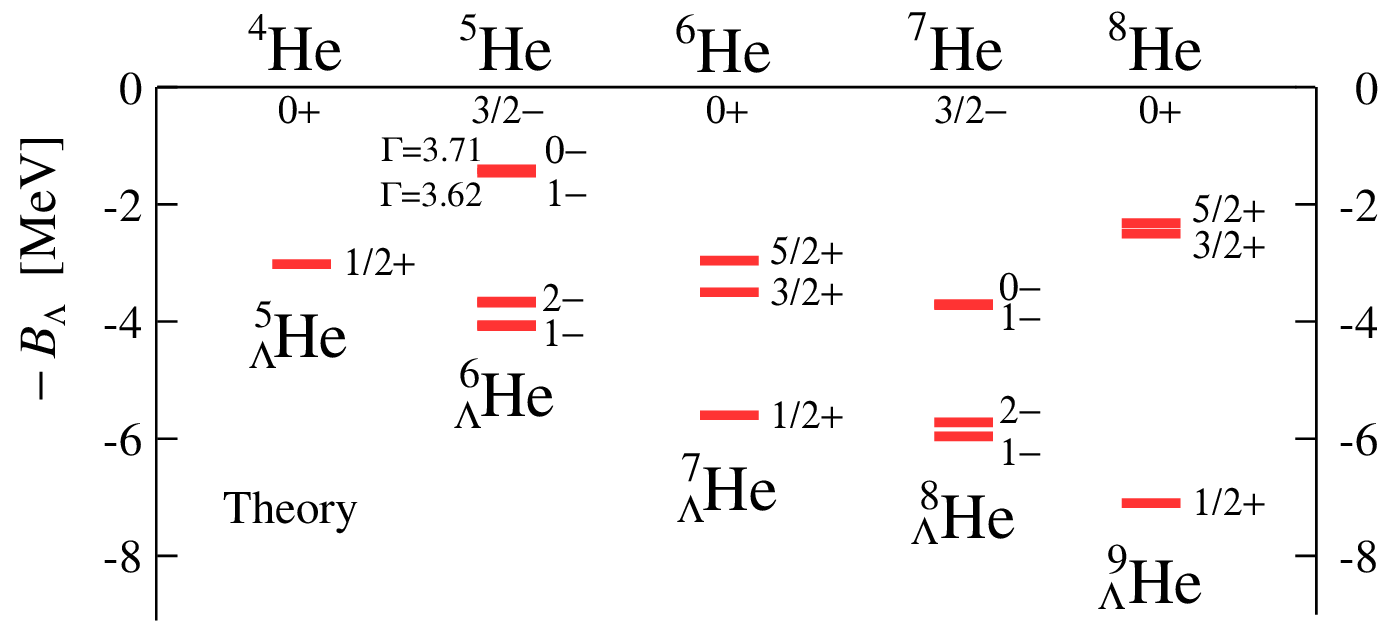}
\caption{The calculated energy spectra  of the He isotope hypenuclei.
 The energies are measured with respect to the $^A$He$ +\Lambda$ ($A=4$,5,6,7,8) thresholds.}
\label{fig:he}
\end{center}
\end{figure}

Next, we discuss on the energy spectra of
$^5_{\Lambda}$He, $^6_{\Lambda}$He,
$^7_{\Lambda}$He, $^8_{\Lambda}$He and $^9_{\Lambda}$He as shown in Fig. 4.
The detailed results for these hypernuclei together with corresponding core
nuclei are listed in Table I.

The $\alpha \Lambda$ potential is adjusted so as to reproduce the
observed binding energy of $^5_{\Lambda}$He.
For the energies of the $A=6$ to 9 $\Lambda$ hypernuclei,
we have no adjustable parameters.
For the ground state of $^6_{\Lambda}$He,
the observed biding energy is $4.18$ MeV with respect to
the $^5{\rm He}+\Lambda$ threshold, which is consistent with
our result, $4.26$ MeV.
This state was considered to have a neutron halo since the observed binding
energy is weakly bound by 0.17 MeV from the lowest threshold,
$^5_{\Lambda}{\rm He}+\Lambda$.
Other theoretical states, $J^\pi =2^-, 1^-_2, 0^-$ states are
resonant states.

$^7_{\Lambda}$He is one of the lightest neutron-rich $\Lambda$ hypenuclei
whose core nucleus is the well-known neutron halo nucleus $^6$He.
Our calculated $B_{\Lambda}$s of the ground state
 and excited states for $J^\pi=1/2^+, 3/2^+$, and $5/2^+$ are 5.51, 3.34, and 2.87 MeV,
respectively. Our ground state energy is in good agreement with
the data. Regarding the excited state, although
it is not clear which excited state, $3/2^+$ or $5/2^+$, is observed,
our averaged energy of two states of $3/2^+$ and $5/2^+$ is not inconsistent with the data.

For $^8_{\Lambda}$He, there have been six events for this events by emulsion data
and then, the observed $B_{\Lambda}$ is $7.16 \pm 0.7$ MeV
\cite{Davis,Gal}, which has large error bar.
Our theoretical calculation in this hypernucleus is
shown in Fig.4.
The core nucleus, the ground state of $^7$He, is a resonant state with a very narrow
decay width. By addition of a $\Lambda$
the ground state, $1^-$ in the resultant hypernucleus $^8_{\Lambda}$He,
becomes a weakly bound state, 0.05 MeV with the respect to
the lowest threshold, $^7_{\Lambda}$He$+n$.
Our theoretical $B_{\Lambda}$ is also listed in Table I, 5.95 MeV, which is
much less binding than observed data.
For $A=8$ hypernuclei, there are other observed $\Lambda$ hypernuclei
such as $^8_{\Lambda}$Li and $^8_{\Lambda}$Be.
These observed $B_{\Lambda}$s are
$6.80 \pm 0.03$ MeV and $6.84 \pm 0.05$ MeV for
$^8_{\Lambda}$Li and $^8_{\Lambda}$Be, respectively.
It should be noted that these two observed $B_{\Lambda}$s are
larger than that of $^8_{\Lambda}$He.
The ground states of core nuclei,  $^7$Li and $^7$Be, are bound while
that of $^7$He is the resonant state.
Generally speaking, by participantion of a $\Lambda$ state, 
$B_{\Lambda}$ for the bound ground state is larger than that for the resonant ground state.
We can see this tendency for $^7_{\Lambda}$He, $^7_{\Lambda}$Li ($B_{\Lambda}^{\rm exp.}=5.58\pm 0.03$ \cite{Davis,Gal}) and
$^7_{\Lambda}$Be ($B_{\Lambda}^{\rm exp.}=5.16\pm 0.08$ \cite{Davis,Gal}).
For $^8_{\Lambda}$He, there have been six events for this events by emulsion data.
Then, we need more data to confirm the binding energy of this hypernucleus.
Other states in $^8_{\Lambda}$He 
 are resonant states. In this system, we obtained resonant states using a
bound state approximation. Thus, the energies are reported without decay widths.

\begin{table*}[t]
\begin{center}
%\begin{minipage}[tbh]{18.0cm}
\caption{ 
Calculated energies of the low-lying states of 
$^6_\Lambda$He, $^7_\Lambda$He, $^8_{\Lambda}$He, and $^9_{\Lambda}$He  
together 
  with those of the corresponding states of $^5$He, $^6$He, $^7$He, and $^8$He,  respectively.
$E$ stands for the total interaction energy among constituent particles.
The energies in parentheses are decay widths for resonant states.
}
\label{tab:halo_energy_rms_A=7}
%\end{minipage}
\begin{tabular}{cccccccccccc}
%\begin{minipage}{0.80\hsize}
\hline \hline  &&&&&&\\[-3mm] 
   & \multicolumn{2}{c}{$^{5}$He($\alpha n$)} & \hspace{5mm}
   & \multicolumn{3}{c}{$^{6}_{\Lambda}$He$(\alpha n\Lambda)$} \\
$J^{\pi}$                   &$3/2^-$        &\quad $1/2^-$      & & $1^-$           & $2^-$         &$1^-$         &$0^-$ \\ \hline %
$E$          (MeV)          &$0.75 (0.59) $ &\quad $2.14(5.84)$ & & $-3.35$         & $-2.94(0.06)$ &$-0.74(3.62)$ &$-0.67(3.71)$ \\
$E^{\rm exp}$(MeV)          &$0.798(0.648)$ &\quad $2.07(5.57)$ & &                 &               &              \\
$B_{\Lambda}$(MeV)          &               &                   & & $4.10$          &   $3.69$      &$1.49$        & $1.42$   \\
$B_{\Lambda}^{\rm exp}$(MeV)&               &                   & & $4.18 \pm 0.10$ &   &        \\
   & \multicolumn{2}{c}{$^{6}$He($\alpha nn)$} & \hspace{5mm}
   & \multicolumn{3}{c}{$^{7}_{\Lambda}$He($\alpha nn\Lambda)$} \\ 
$J^{\pi}$          &  $0^+$  &      $2^+$          & & $1/2^+$ & $3/2^+$ &$5/2^+$ \\ \hline %
$E$          (MeV) &$-0.975$ &\quad $0.879(0.132)$ & & $-6.48$ & $-4.37$ &$-3.84$ \\
$E^{\rm exp}$(MeV) &$-0.975$ &\quad $0.822(0.113)$ & &         &         &        \\
$B_{\Lambda}$(MeV) &         &                     & & $5.51$  & $3.34$  &$2.87$  \\
$B_{\Lambda}^{\rm exp}$  (MeV)	&  & &   &$5.55\pm 0.10({\rm stat.})$
 & $3.65 \pm 0.20({\rm stat.})$
  &        \\
	&  & &   &
$\pm 0.11({\rm sys.})$ & $\pm 0.11({\rm sys.})$
  &        \\
   & \multicolumn{2}{c}{$^{7}$He$(\alpha nnn)$} & \hspace{5mm}
   & \multicolumn{3}{c}{$^{8}_{\Lambda}$He $(\alpha nnn \Lambda)$} \\ 
$J^{\pi}$                     &$3/2^-$        & $1/2^-$    & & $1^-$   & $2^-$  &$1^-$   &$0^-$\\ \hline %
$E$	     (MeV)            &$-0.58(0.048)$ & 0.42(2.77) & & $-6.53$ &$-6.33$ &$-4.31$ &$-4.27$ \\
$E^{\rm exp}$(MeV)            &$-0.53(0.15) $ & 0.37(1.0)  & &         &        &         \\
$B_{\Lambda}$(MeV)            &               &            & & $5.95$  &$5.75$  &$3.73$  &$3.69$  \\
$B_{\Lambda}^{\rm exp}$ (MeV) &               &            & & $7.17 \pm 0.7$ &  &        \\
   & \multicolumn{2}{c}{$^{8}$He($\alpha nnnn)$}                  & \hspace{5mm}
   & \multicolumn{3}{c}{$^{9}_{\Lambda}$He($\alpha nnnn\Lambda)$} \\ 
$J^{\pi}$          &$0^+$    &      $2^+$        &   & $1/2^+$  & $3/2^+$       &$5/2^+$       \\ \hline %
$E$	     (MeV) & $-3.21$ &\quad $0.32(0.66)$ &   & $-10.29$ & $-5.69(<0.1)$ &$-5.53(<0.1)$ \\
$E^{\rm exp}$(MeV) & $-3.11$ &\quad $0.43(0.89)$ &   &          &               &              \\
$B_{\Lambda}$(MeV) &         &                   &   & $7.09$   & $2.49$        &$2.33$        \\
\hline \hline
%\end{minipage}
\end{tabular}
\end{center}
\end{table*}     
%%%%%%%%%%%%%%%%%%%%%%%%%%%%%%%%%%%%%%%%%%%%%%%%%%%%%%%%%%%%%%%%%%%%%%%%%%%%%%%%%%%%%%%
% 7He(1/2-) exp taken from F. Skaza et al., Physical Review C 73, 044301 (2006)
% 8He(2+)   exp taken from M. Holl et al., Physics Letters B 822, 136710 (2021)

Finally, let us consider the energy spectra of $^9_{\Lambda}$He.
The core nucleus, $^8$He has one bound state and several resonant excited states.
As mentioned before,
our calculated energy levels of this core nucleus are in good agreement with
the observed data.
With the addition of a $\Lambda$ particle, the ground state of $^9_{\Lambda}$He
becomes deeply bound, having a $\Lambda$ separation energy of
$B_\Lambda= 7.09$ MeV.
The first excited state of $^8$He, a $2^+$ state, is a resonant state
at 0.32 MeV with $\Gamma =0.66$ MeV.  Even if the $\Lambda$ 
particle is injected into this state, the $\Lambda N$ attraction is not
enough to make this state bound. The resulting states,
$3/2^-$ and $5/2^-$ states, are narrow resonant states (less than 100 keV widths).
Thus, the relative energy  ($\approx 4.7$ MeV)  between the $1/2^+$ and the 
average of $3/2^+$ and $5/2^+$ in $^9_{\Lambda}$He is larger than the energy 
 ($\approx 3.5$ MeV) between $0^+$ and $2^+$ in $^8$He. 
Therefore, we have one deeply bound state for $^9_{\Lambda}$He.
Currently, it is planned to produce this hypernucleus at J-PARC using the $(\pi^-, K^+)$ reaction
on a  $^9$Be target with a resolution of about 3 MeV.
According to our calculation, the relative energy between the ground state and
the excited state is 4.6 MeV. 
Then, it would be possible to separate the ground state and the first excited state.
Experimentally, there was a pioneering experiment, a
double-charge-exchange reaction $(\pi^-, K^+)$ on a $^{10}$B target to
produce $^{10}_{\Lambda}$Li, although it was difficult to obtain significant discrete peaks
due to the small cross section  \cite{Saha2005}.
With this experience, it should be possible to produce $^9_{\Lambda}$He for the
first time at J-PARC. This hypernucleus, $^9_{\Lambda}$He,
would be a most exotic system.

\section{Summary}
Motivated by the recently observed neutron-rich $\Lambda$ hypernucleus $^7_{\Lambda}$He
and future plans to produce $^9_{\Lambda}$He, we have studied the
series of He isotope $\Lambda$ hypernuclei, from
$^6_{\Lambda}$He to $^9_{\Lambda}$He within the framework of an $\alpha +Xn+\Lambda$ ($X=1$--4) cluster model.
The interactions employed between constituent particles are based upon 
the observed data. The calculated energy spectra of the core nuclei with $A=5$ to 8 are
consistent with the published data.
 For the resonant states of $^5$He, $^6$He, and $^7$He, we employ the complex scaling method to obtain
energies and decay widths, and our results are consistent with the
observed resonant states.
Our results for $^6_{\Lambda}$He and $^7_{\Lambda}$He are in good agreement with the
available data.
We also compared our results of  $^8_{\Lambda}$He and data by emulsion data.
We see that theoretical binding energy is less bound with the data.
However, due to large error bar and small number of events for it,
we need more precise data to discuss on this hypernucleus.
Within the present theoretical framework, we predict a very weakly bound ground
state, $1^-$ (50 keV bound with respect to the $^7_{\Lambda}$He$+n$ threshold)
and three resonant states.
We have also calculated the energy levels of $^9_{\Lambda}$He.
We obtain one deeply bound state, a $1/2^+$ state, and two narrow resonant states,
$3/2^+$ and $5/2^+$.
It is planned to produce the $\Lambda$ hypernucleus $^9_{\Lambda}$He at J-PARC
utilizing the $(\pi^-, K^+$) reaction on a $^9$Be target.
Because the calculated ground state binding energy is about 4 MeV below the lowest threshold,
$^7_{\Lambda}$He$+n$, and there are no states around the threshold, it should be possible
to identify the state.
If $^9_{\Lambda}$He is observed at J-PARC, its binding energy would be a useful
constraint, along with those of the lighter $\Lambda$ hypernuclei, in modeling 
$\Lambda N$-$\Sigma N$ coupling.

%
%-------------  Acknowledgment ----------
%
\section*{Acknowledgments}
The authors thank Dr.\ B.\ F.\ Gibson for helpful discussions.
This work was supported by a Grant-in-Aid for
Scientific Research on Innovative Areas, Grant No. JP18H05407
and JSPS KAKENHI Grants No. JP20H00155, No. JP18K03660, and No. JP22K03643.
Numerical calculations were partly achieved through the use of SQUID at the Cybermedia Center, Osaka University.


\begin{thebibliography}{99}

\bibitem{Nakamura}
S. N. Nakamura {\it et al.} (HKS(Jlab E01-011) Collaboration),
Phys. Rev. Lett. {\bf 110}, 012502 (2013).
 
\bibitem{Gogami}
T. Gogami {\it et al.},  (HKS(Jlab E05-115) Collaboration),
Phys. Rev. C {\bf 94}, 021302 (R) (2016).


\bibitem{Hiyama96} E. Hiyama, M. Kamimura, T. Motoba,
T. Yamada, and Y. Yamamoto, Phys. Rev. C{\bf 53}, 2075 (1996).


\bibitem{Hiyama2009} E. Hiyama, Y. Yamamoto, T. Motoba,
and M. Kamimura, Phys. Rev. C {\bf 80}, 054321 (2009).


\bibitem{Hiyama2015}
E. Hiyama, M. Isaka, M. Kamimura, T. Myo,  and T. Motoba,
Phys. Rev. C {\bf 91}, 054316 (2015).

\bibitem{Hiyama2001} E. Hiyama, M. Kamimura,
T. Motoba, T. Yamada, and Y. Yamamoto, 
Phys. Rev. C{\bf 65}, 011301(R) (2001).

\bibitem{Nogga} A. Nogga, H. Kamada, and W. Gloeckle,
Phys. Rev. Lett.  {\bf 88}, 172501 (2002).


\bibitem{Nemura} H. Nemura, Y. Akaishi, and Y. Suzuki,
Phys. Rev. Lett. {\bf 89}, 142504 (2002).


\bibitem{wirth}  R. Wirth and R. Roth, 
Phys. Lett. B {\bf 779}, 336 (2018).

\bibitem{Machleidt}D. R. Enterm and R. Machleidt,
Phys. Rev. C {\bf 68}, 041001 (2003).

\bibitem{Navratil} P. Navr\'{a}til, Few-body sys. {\bf 41}, 117 (2007).

\bibitem{Haidenbauer} J. Haidenbauer, S. Petschauer, N. Kaiser, U.-G. Mei${\ss}$ner, A. Nogga,
and W. Weise, Nucl. Phys. {\bf A} 915, 24 (2013).

\bibitem{myo21}  T. Myo, M. Odsuren, K. Kat\=o, Phys. Rev.\ C {\bf 104}, 044306 (2021).  % 8He, 8C

% myo reference
\def\JL#1#2#3#4{ {{\rm #1}} \textbf{#2}, #4 (#3)}  % Physical Review
%\def\JL#1#2#3#4{ {{\rm #1}} #2 (#3) #4}  % Elsevier
\nc{\PR}[3]     {\JL{Phys. Rev.}{#1}{#2}{#3}}
\nc{\PRC}[3]    {\JL{Phys. Rev.~C}{#1}{#2}{#3}}
\nc{\PRA}[3]    {\JL{Phys. Rev.~A}{#1}{#2}{#3}}
\nc{\PRL}[3]    {\JL{Phys. Rev. Lett.}{#1}{#2}{#3}}
\nc{\NP}[3]     {\JL{Nucl. Phys.}{#1}{#2}{#3}}
\nc{\NPA}[3]    {\JL{Nucl. Phys.}{A#1}{#2}{#3}}
\nc{\PL}[3]     {\JL{Phys. Lett.}{#1}{#2}{#3}}
\nc{\PLB}[3]    {\JL{Phys. Lett.~B}{#1}{#2}{#3}}
\nc{\PTP}[3]    {\JL{Prog. Theor. Phys.}{#1}{#2}{#3}}
\nc{\PTPS}[3]   {\JL{Prog. Theor. Phys. Suppl.}{#1}{#2}{#3}}
\nc{\PRep}[3]   {\JL{Phys. Rep.}{#1}{#2}{#3}}
\nc{\JP}[3]     {\JL{J. of Phys.}{#1}{#2}{#3}}
\nc{\PPNP}[3]   {\JL{Prog. Part. Nucl. Phys.}{#1}{#2}{#3}}
\nc{\PTEP}[3]   {\JL{Prog. Theor. Exp. Phys.}{#1}{#2}{#3}}
\nc{\andvol}[3] {{\it ibid.}\JL{}{#1}{#2}{#3}}


\bibitem{suzuki88}   Y. Suzuki and K. Ikeda, \PRC{38}{1988}{410}.  % COSM model

\bibitem{masui06}    H. Masui, K. Kat\=o,~K. Ikeda, \PRC{73}{2006}{034318}. % COSM He isotopes

\bibitem{myo14}  T. Myo, Y. Kikuchi, H. Masui, K. Kat\=o, Prog. Part. Nucl. Phys. {\bf 79}, 1 (2014).  % Review

\bibitem{myo22}  T. Myo, K. Kat\=o, Phys.\ Rev.\ C {\bf 106}, L021302 (2022). % 8He, soft dipole mode, letter




\bibitem{kanada79} H.\ Kanada, T.\ Kaneko, S.\ Nagata, and M.\ Nomoto, Prog.\ Theor.\ Phys.\ {\bf 61}, 1327 (1979). % alpha-n potential
\bibitem{tang78} Y. C. Tang, M. LeMere and D. R. Thompson, Phys. Rep. {\bf 47}, 167 (1978).  % n-n potential

\bibitem{motoba85} T. Motoba, H. Band\=o, K. Ikeda, and T. Yamada, Prog.\ Theor.\ Phys.\ Suppl. {\bf 81}, 42 (1985) % hyper review
\bibitem{hiyama97}   E. Hiyama, M. Kamimura, T. Motoba, T. Yamada, and Y. Yamamoto, \PTP{97}{1997}{887}.  % Hyper four-body

\bibitem{Hiyama03} E.~Hiyama, Y.~Kino and M.~Kamimura, 
Prog.\ Part.\ Nucl.\ Phys.\ {\bf 51}, 223 (2003).

\bibitem{myo10}      T. Myo, R. Ando and K. Kat\=o, \PLB{691}{2010}{150}. % 8He




\bibitem{aoyama06}   S. Aoyama, T. Myo, K. Kat\=o, K. Ikeda, \PTP{116}{2006}{1}. % review  CSM

\bibitem{myo20}      T. Myo, K. Kat\=o, \PTEP{2020}{2020}{12A101}.  % Review CSM

\bibitem{ho83}       Y. K. Ho, \PRep{99}{1983}{1}. % CSM review 

\bibitem{moiseyev98} N. Moiseyev, \PRep{302}{1998}{211}. % CSM review 




\bibitem{ABC}        J. Aguilar and J.M.Combes, \JL{Commun. Math. Phys.}{22}{1971}{269}.
	             E. Balslev and J.M. Combes, \JL{Commun. Math. Phys.}{22}{1971}{280}.

\bibitem{berggren68} T. Berggren,~\NP{A109}{1968}{265}. % bi-orthogonal

\bibitem{Tilley} D.R. Tilley, C.M. Cheves, J.L. Godwin, 
G.M. Hale, H.M. Hofmann, J.H. Kelley, C.G. Sheu, H.R. Weller,  Nucl. Phys. A 708, 3 (2002).

\bibitem{Tilley-he}
D. R. Tilley, J. H. Kelley, J. L. Godwin, D. J. Millener, J. E. Purcell, C. G. Sheu, and H.R. Weller, Nucl. Phys. A745, 155 (2004). 

\bibitem{Skaza} F. Skaza et al., Phys. Rev. C73, 044301 (2006), and the references therein. 

\bibitem{Golovkov} M. S. Golovkov et al., Phys. Lett. B 672, 22 (2009), and the references therein.

\bibitem{Holl} M. Holl et al., Phys. Lett. B, 822, 136710 (2021).

I. Tanihata, D. Hirata 1, 

\bibitem{tanihata92} I. Tanihata, D. Hirata, T. Kobayashi, S. Shimoura, K. Sugimoto, H. Toki,~\PLB{289}{1992}{261}.  % 8He radius
\bibitem{alkazov97}  G. D. Alkhazov {\it et al.}, \PRL{78}{1997}{2313}. % 8He radius
\bibitem{kiselev05}  O. A. Kiselev {\it et al.},  \JL{Eur. Phys. J. A25}{Suppl.1}{215}{2005}. % 8He radius
\bibitem{mueller07}  P. Mueller {\it et al.}, \PRL{99}{2007}{252501}. % 8He charge radius
\bibitem{brodeur12}  M. Brodeur {\it et al.}, \PRL{108}{2012}{052504}. % 8He charge


\bibitem{Saha2005} P.K. Saha {\it et al.}, Phys. Rev. Lett. {\bf 94}, 052502 (2005).


\bibitem{Davis} D. H. Davis, and J. Pniewski,
Contemp. Phys. {\bf 27}, 91 (1986).

\bibitem{Gal} A. Gal, E. V. Hungerford, and D. J. Millener,
Rev. Mod. Phys. {\bf 88}, 035004 (2016).
 
\end{thebibliography}
\end{document}